%% file: main.tex
\documentclass[conference]{IEEEconf}
\input epsf
\usepackage{graphicx}
\usepackage{amsmath}
\usepackage{xcolor, soul}
\usepackage{listings}

\definecolor{codebackground}{rgb}{0.95, 0.95, 0.92}
\let\origthelstnumber\thelstnumber
\makeatletter
\newcommand*\stopnumber
{
    \lst@AddToHook{OnNewLine}
    {
        \let\thelstnumber\relax
        \advance\c@lstnumber-\@ne\relax
    }
}
\newcommand*\startnumber[1]
{
    \setcounter{lstnumber}{\numexpr#1-1\relax}
    \lst@AddToHook{OnNewLine}
    {
        \let\thelstnumber\origthelstnumber
        \refstepcounter{lstnumber}
    }
}
\makeatother

\usepackage{caption}
\usepackage{xurl}

\begin{document}

% paper title
%\title{Submission Format for IPVC-CyberSec21 (Title in 24-point Times font)}
% If the \LARGE is deleted, the title font defaults to  24-point.
% Actually, 
% the \LARGE sets the title at 17 pt, which is close enough to 18-point.
%+++++++++++++++++++++++++++++++++++++++++++
\title{\textbf{\Large Simulation of Sensor Spoofing Attacks on Unmanned Aerial Vehicles Using the Gazebo Simulator\\}}

\author{Irdin Pekaric$^{*}$, David Arnold$^{}$, Michael Felderer$^{}$\\
	\normalsize University of Innsbruck, Innsbruck, Austria\\
	\normalsize irdin.pekaric@uibk.ac.at, david.arnold@student.uibk.ac.at, michael.felderer@uibk.ac.at\\
	\normalsize *corresponding author
}

%+++++++++++++++++++++++++++++++++++++++++++

% use only for invited papers
%\specialpapernotice{(Invited Paper)}

% make the title area
\maketitle
\begin{abstract}

Conducting safety simulations in various simulators, such as the Gazebo simulator, became a very popular means of testing vehicles against potential safety risks (i.e. crashes). However, this was not the case with security testing. Performing security testing in a simulator is very difficult because security attacks are performed on a different abstraction level. In addition, the attacks themselves are becoming more sophisticated, which directly contributes to the difficulty of executing them in a simulator. In this paper, we attempt to tackle the aforementioned gap by investigating possible attacks that can be simulated, and then performing their simulations. The presented approach shows that attacks targeting the LiDAR and GPS components of unmanned aerial vehicles can be simulated. This is achieved by exploiting vulnerabilities of the ROS and MAVLink protocol and injecting malicious processes into an application. As a result, messages with arbitrary values can be spoofed to the corresponding topics, which allows attackers to update relevant parameters and cause a potential crash of a vehicle. This was tested in multiple scenarios, thereby proving that it is indeed possible to simulate certain attack types, such as spoofing and jamming.

\end{abstract}
\IEEEoverridecommandlockouts

\IEEEpeerreviewmaketitle

\section{Introduction}
\label{sec:intro}
\input{Chapters/Introduction}

\section{Background}
\label{sec:backg}
\input{Chapters/Background}

\section{Related Work}
\label{sec:rw}
\input{Chapters/Related}

\section{Simulation Setup}
\label{sec:exp}
\input{Chapters/Experiment}

\section{Attack Simulations}
\label{sec:attsim}
\input{Chapters/AttackSimulations}

\section{Evaluation}
\label{sec:eval}
\input{Chapters/Evaluation}

\section{Discussion}
\label{sec:diss}
\input{Chapters/Discussion}

\section{Conclusion}
\label{sec:concl}
\input{Chapters/Conclusion}

\section*{Acknowledgment}
This work was partially supported by the Austrian Science Fund (FWF): I 4701-N.
\bibliographystyle{IEEEtran}
\bibliography{references}

\end{document}

%% file: Chapters/Introduction.tex
Unmanned Aerial Vehicles (UAVs) represent one of the fastest growing types of vehicles worldwide. They are used in various domains, such as military, medical, and transportation. Every year, the number of drones is rapidly increasing and it is estimated that this number will reach 400.000 by the end of the year 2050 \cite{doi/10.2829/085259}. 

UAVs heavily rely on different types of sensors in order to gain some insights into the surrounding environment, weather, and telemetry. This makes them prone to security risks that can result in devastating safety hazards \cite{ramos2021security}. In order to prevent this, each type of UAV needs to be rigorously tested. For example, this can be done by applying penetration testing and dynamic analysis techniques \cite{veerappan2021drat}. However, this is not always that simple because it can result in physical damage to a system.

As a result, various simulators such as the Gazebo simulator \cite{gazebo} were developed, which allow vehicles to be tested virtually in safe conditions. In past years, simulators were used heavily for the purpose of safety testing. However, security testing was rarely applied because it is very difficult to conduct such testing in a simulator. This is due to the fact that security testing is done on a different abstraction level compared to safety testing \cite{witte2022towards}.

In this paper, we demonstrate that security testing in a simulator is possible to a certain degree. This is achieved by simulating attacks on the perception-based components of UAVs. More specifically, we identify existing vulnerabilities of technologies utilized by UAVs and exploit them in order to experimentally conduct attack simulations. Attacks that we address target the vehicle using their sensors, such as the LiDAR. The presented approach provides a good basis for conducting further security simulations and observing possible causes of specific attacks.

The remainder of this paper is structured as follows: Section \ref{sec:backg} provides background information on the Gazebo simulator and the simulated attacks as well as exploited vulnerabilities. Section \ref{sec:rw} discusses the related work on security attack simulations and sensor attacks. 
Section \ref{sec:exp} demonstrates the implementation of obstacle detection and avoidance with LiDAR and automatic waypoint navigation that were used to conduct simulations.
Section \ref{sec:attsim} presents the simulations themselves, as well as the implemented attack scenarios. Section \ref{sec:eval} provides an evaluation of the proposed approach. Section \ref{sec:diss} discusses key findings and offers a comparison between simulations and real-world evaluations. Finally, Section \ref{sec:concl} concludes the paper and provides an outlook on future work. 

%% file: Chapters/Background.tex
In this section, the background information on the Gazebo simulator, Robot Operating System (ROS), addressed sensor components, and Guidance Navigation and Control (GNC) API are presented. These are all utilized in the proposed approach in order to conduct attack simulations.

\subsection{Gazebo and ROS}

The Gazebo framework is one of the few that attempts to recreate realistic scenarios for robots and not just for humans \cite{qian2014manipulation}. The simulations are conducted in both indoor and outdoor three-dimensional spaces with an aim to accurately represent any environments and objects that a robot may encounter. The gazebo has both standard and native interfaces. Its clients can access all the needed data via shared memory \cite{takaya2016simulation}. Once a robotic object is created in a simulator, it is possible to associate it with one or more controllers. Whenever any data is generated as a result of controller interaction, this data is then stored in shared memory using Gazebo interfaces or so-called Ifaces. By utilizing the Ifaces, inter-process communication can be achieved from shared memory \cite{takaya2016simulation}.

ROS includes a set of tools and libraries that allow for the easier development of robotic applications. Due to the recent rapid development of robots, it is necessary to build a common set of software that can be reused among different robotic models. For example, this can include navigation, motion planning, or mapping applications \cite{o2014gentle}. ROS and Gazebo are integrated together through the gazebo-ros package, which allows bidirectional communication \cite{raje2020evaluation}. For example, the sensor data can flow from the Gazebo to ROS, while actuator commands can be sent back from ROS to Gazebo.

\subsection{Addressed Sensor Components}

The two most applied sensor components of a UAV include LiDAR and Global Positioning System (GPS) sensors. 

\begin{figure}%[h]
    \begin{center}
        \includegraphics[width=0.8\columnwidth]{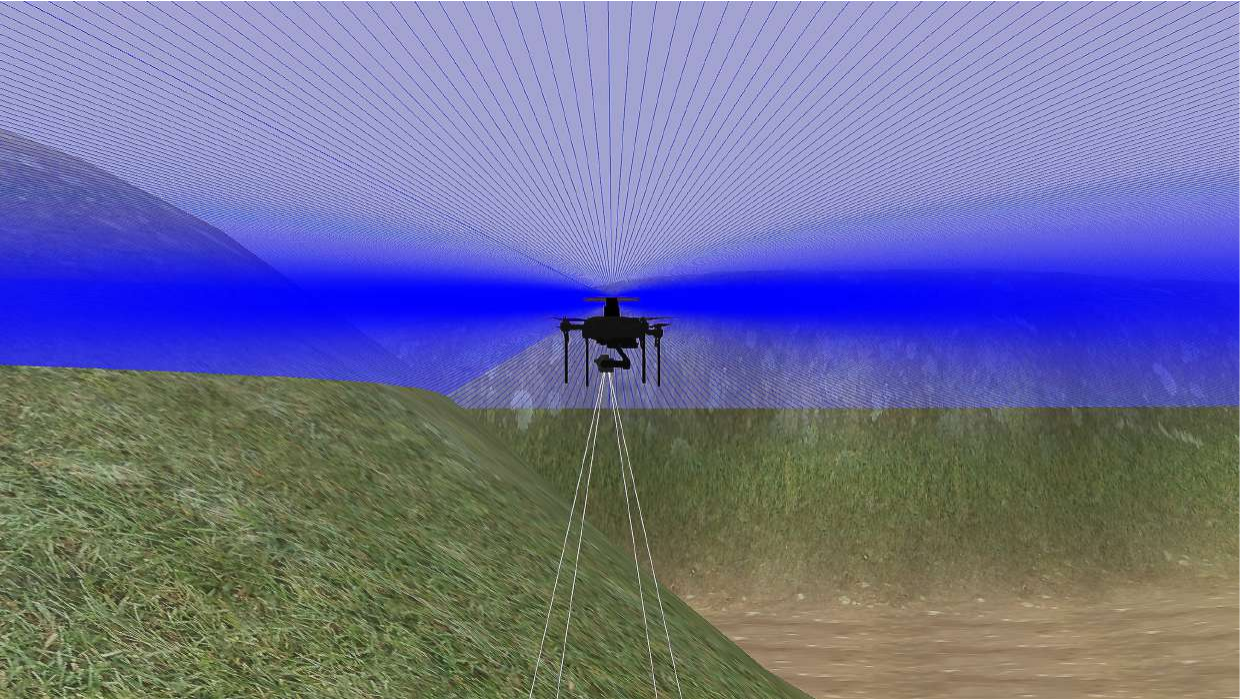}
    \end{center}
    \caption{The iris model with the LiDAR sensor.}
    \label{fig:Iris}
\end{figure}

LiDAR stands for Light Detection and Ranging \cite{reutebuch2005light}. This means that its task is to map out the surrounding environment, as well as any other objects and vehicles that are located close by. This is achieved by emitting an incoherent optical signal, which is modulated by a cosine-shaped signal. In most cases, this signal operates at a frequency within the non-visible spectrum that is close to the infrared spectral range at around 780 nanometres \cite{khan2021comparative}. For the implementation of attack scenarios, we use a two-dimensional LiDAR sensor developed by the Hokuyo \footnote{https://hokuyo-usa.com/products/lidar-obstacle-detection/ [Accessed May-2022]} manufacturer, which is provided by the Guidance, Navigation, and Control (GNC) API v5.1.
The sensors developed by this manufacturer have a maximum field of view between 180 and 270 degrees. However, the default field-of-view of the API of the simulated LiDAR is 360 degrees. Considering that the aim of this research is not to investigate various attacks on specific hardware, the range was kept at 360 degrees. Figure \ref{fig:Iris} shows the iris model with the LiDAR sensor
attached to it provided by the GNC API. 

GPS is the first Global Navigation Satellite
System (GNSS) to come into function \cite{wells1987guide}. It is the most widely used system for navigation provided using satellites. A set of countries that include the USA (GPS- 31 satellites), Russia (GLONASS- 23 satellites), China (Bei-Dou- 44 satellites), and EU (Galileo- 24 satellites) provide the world with GPS tracking using their own systems \cite{tiesinga2021codata}. Based on the research by Rozenbeek et al. \cite{Evaluation2020Rozenbeek}, a combination of both GPS and GLONASS is mainly used for GNSS navigation of commercially available UAVs. Consequently, such a combination of two different navigational systems leads to an improvement in positioning accuracy. 

\subsection{Guidance Navigation and Control API}
\label{sec:gnc}

The GNC API \cite{IntelligentGnc2022Johnson}, provides a comprehensive software collection with the aim of enabling developers to make the development of intelligent aircraft more accessible by sharing various libraries and plugins within the community. More specifically, the GNC API focuses on functions that are helpful for designing guidance programs for UAVs. It is worth mentioning that the GNC community provides not only the API but also a comprehensive collection of example simulations \cite{IntelligentSim2022Johnson}. The initial version of the GNC API and the version primarily used for this work were developed in C++ programming language. In comparison to the Python version, a solution implemented and pre-compiled in C++ language will result in better performance \cite{Programming2014Bartonicek}. Given that performance is a highly important
aspect when conducting simulations, all the simulations that are a part of this research were created using the C++-based version of the GNC API. On the other hand, the generation of the attacks described in Section \ref{sec:attsim} is based on Python.
The most significant functions of the API that were used during the implementation of the simulations presented in this research include \textit{wait4start()}, \textit{initialize\_local\_frame()}, \textit{takeoff(float takeoff\_alt)}, \textit{set\_destination(float x, float y, float z, float psi)}, \textit{check\_waypoint\_reached(float pos, float heading)}, and \textit{land()}.

%% file: Chapters/Related.tex
The related work on jamming and spoofing attacks, security simulations utilizing ROS and Gazebo, and vulnerabilities of the MAVLink protocol are discussed as follows. In addition, we also pinpoint the differences between the related work and contributions of this paper.

\subsection{Jamming and Spoofing Attacks}

In regards to UAVs, jamming and spoofing attacks most often target GPS and LiDAR components. 

Kerns et al. \cite{kerns2014unmanned} showed that it is possible to take control of a drone by applying GPS spoofing methods. To achieve this, an attacker used authentic signals sent by a UAV, which were used to generate counterfeit signals. These signals were then sent back to the drone, which would shift the position of a specific object by a certain amount. This was proven in an experiment in which the odometry information of a hovering UAV was spoofed. As a result, the drone was shown as if it was moving upward. However, the correct movement of a drone was downwards. Kerns et al. \cite{kerns2014unmanned} also differentiated between overt and covert attacks. In regards to covert attacks, the detection methods of a defending system were considered, such as Y/N monitoring. It is triggered when the received power significantly exceeds the one under quiescent conditions. However, a large majority of systems do not have any detection methods implemented due to them being too costly or hard to use. For the overt attack to be successful, only a certain signal difference in regard to the spoofing power is needed. 

The work of Kerns et al. \cite{kerns2014unmanned} was further extended by Eldosouky et al. \cite{eldosouky2019drones} who introduced the mathematical framework that protects UAVs against spoofing attacks. It allows a detailed analysis of GPS-related attacks. In addition, Eldosouky et al. \cite{eldosouky2019drones} demonstrated an attack that targets the odometry of a UAV and makes it change the correct flight path with another malicious path. The authors emphasized that it is very significant to always choose the shortest flight path between the current location and the location imposed by an attacker to prevent any issues created by UAV's guidance logic. The shortest path is calculated according to the euclidian distance.

Gaspar et al. \cite{gaspar2020capture} demonstrated an approach that makes it possible to recreate authentic signals of a real system. This was achieved with a functioning simulator provided by a UAV manufacturer, wherein National Marine Electronics Association (NMEA) messages were generated. The message included a fixed GPS location, which could be adjusted by an adversary by creating a malicious message. This was tested experimentally by spoofing GPS parameters and taking control of a drone. As a result, it was proven that the aforementioned procedure could be used by an attacker to target UAVs.

Petit et al. \cite{petit2015remote} conducted an experimental evaluation of the effects of jamming and spoofing attacks on LiDAR sensors as part of the test set. In their work, they captured and analyzed the properties of an authentic signal. Furthermore, they used an enhanced relaying attack, wherein the original signal is captured and sent to the attacker's destination. They were able to prove that it was possible to spoof a signal off a wall at different distances. For example, if the wall is only five meters away from the vehicle, an attacker could make it seem that the distance is 50 meters. According to all presented attacks in LiDAR sensors, it can be concluded that this component is vulnerable to spoofing attacks that utilize existing, but also non-existing objects.

The work of Petit et al. \cite{petit2015remote} was further extended by Cao et al. \cite{cao2019adversarial} who addressed spoofing attacks against the LiDAR component of modern autonomous vehicles. The goal of the attacks was to spoof various objects that were located in front of the target vehicle. Compared to the work of Petit et al. \cite{petit2015remote}, the temporal delay was significantly reduced, which made it possible to reduce the shortest distance to an obstacle to two meters only. However, modern autonomous vehicles have defensive capabilities against LiDAR spoofing attacks that are based on machine learning algorithms known as spoofing detection techniques. This is because the exact location of spoofed points does not correspond to a real object. Thus, it was necessary to not only manipulate the distance, but also the altitude of the spoofed signals. This led to the conclusion that attacks on the LiDAR component of a complex system need to be executed as covert attacks. 

\subsection{Security Simulations with ROS and Gazebo Simulator}

Spina et al. \cite{spina2021integrating} indicated that both ROS and Gazebo do not consider security aspects. This means that developers and security testers cannot conduct security testing unless they make significant modifications to a simulator itself. The authors implemented different security scenarios together with possible mitigation techniques in order to evaluate existing security issues. An example was an attack that targets the integrity and confidentiality of messages that exchange data between different nodes in a network. As a result, all other robotic models can read data of any node in the network and perform manipulations in order to disrupt normal behavior. In order to prevent this, the authors implemented encryption mechanisms, which allow the encryption and decryption of messages using a symmetric key. As a result, an attacker cannot use a malicious robot or vehicle to read other private messages that were intended for other nodes.

Dieber et al. \cite{dieber2017security} also noted the importance of implementing security mechanisms in the ROS framework. This was strictly related to the exchange of messages in ROS applications. Dieber et al. \cite{dieber2017security} experimentally showed that a malicious process can gain access to the arbitrary data of an application. This could be achieved without any authorization requirements. In addition, an attack was presented that demonstrates how an adversary can publish fake messages with spoofed data to a running ROS application. Based on the attack, it was deduced was also possible to manipulate the data of various processes purposely by an adversary.

\subsection{Vulnerabilities of the MAVLink Protocol}

Koubâa et al. \cite{koubaa2019micro} analyzed various vulnerabilities of the MAVLink protocol and threats that can result from exploiting these vulnerabilities. The threats were classified into different categories, such as the ones that affect confidentiality, integrity, and availability. For example, jamming attacks can be used to target protocol packets, which can result in an interruption between the ground control station (GCS) and the model. Consequently, this could lead to a model being placed into the lost-link state. 

Another type of attack that was investigated was the flooding attack. This type of attack could target the communication between the GCS and a model causing an overload and stopping some core services. In the experiment conducted by Koubâa et al. \cite{koubaa2019micro}, the attack caused the model to crash. Furthermore, in another attack example, the attacker was able to intercept authentic packets sent by the GCS and modify them in order to target a UAV. This caused an inability to interpret the authentic packets, thereby resulting in a model being uncontrollable. Moreover, Koubâa et al. \cite{koubaa2019micro} applied a manipulation tool, which allowed the spoofing of packets containing wrong position-related information. As a result, it was possible to mislead the GCS into assuming that the model is at a specific (faulty) position.

Kwon et al. \cite{kwon2018empirical} presented several attacks that could be accomplished due to the vulnerabilities identified in the MAVLink protocol. These were carried out with the assumption that an adversary has an access to the local network. This means that the lack of encryption in the protocol can be easily exploited. Aside from the application of flooding attacks, the authors exploited the waypoint functionality of the protocol. This is initiated when a specific packet is received and causes the deletion of all the existing information regarding planned missions and waypoints that relate to these missions. For this reason, the system was put in a pending state where it awaits further instructions. In order to execute the aforementioned procedure, Kwon et al. \cite{kwon2018empirical} applied the packet injection attack, in which the attacker sent packets to signal the start of a mission. This cleared the memory and caused the remainder of the mission to not be carried out. Due to the functionalities of the MAVLink protocol, the waypoints were never re-sent, which made the UAV hover in the air waiting for any new commands. 

\subsection{Beyond State-of-the-Art}

The aforementioned related work that addresses jamming and spoofing attacks on UAVs demonstrated that it is feasible for adversaries to target and affect the movement of a vehicle. This was specifically shown in studies by Petit et al. \cite{petit2015remote} and Cao et al. \cite{cao2019adversarial}, wherein the LiDAR component was targeted in order to deceive a system regarding the distance to surrounding objects. In order to prevent this, extensive security testing needs to be performed. However, this process can be time-consuming because various scenarios that include multiple surrounding objects, environments, and weather conditions need to be evaluated. In addition, this can also create enormous financial costs because of possible damage to a test vehicle. Thus, it is of utmost significance to find an alternative way to perform testing, such as in a simulator. This is achieved in the proposed approach, wherein the possibilities to perform such experimental investigations based on simulations are investigated. More specifically, attacks on GPS and LiDAR components are simulated in the Gazebo simulator.

%% file: Chapters/Experiment.tex
In this section, the obstacle detection and avoidance with LiDAR and automatic waypoint navigation are discussed. These represent the key concepts upon which attack simulations were built. 

\subsection{Obstacle Detection and Avoidance with LiDAR}

As UAVs are able to move relatively unrestricted through their environment, the precise knowledge of the environment is usually unknown before the
flight. In addition, the environment does not always contain static objects, and their positions may change during the flight (e.g. due to people, animals, or even other UAVs). In order to achieve the goal of a collision-free flight in path planning, local obstacle handling has to be applied, which involves both the detection and avoidance of various objects \cite{Obstacle2005Ribeiro}. Obstacle avoidance can be defined as a set of methodologies that shape the path of the robot to overcome unforeseen
obstacles \cite{Obstacle2005Ribeiro}.
A frequently used algorithm for the application of real-time obstacle avoidance is the method of artificial potential fields. The underlying philosophy of this method describes the environment within which the certain model travels as a field of forces.
Compared to an energy field, the field consists of attractive (positive charge) and repulsive (negative charge) forces. The attractive pole within the field is generated by the target position, whereas the obstacles in the field are responsible for generating the repulsive forces \cite{Realtime1995Khatib}. The analogy further implies that a model is considered to be a particle in the field, which moves immersed in the potential field with a desire to reach the target position. In the process, the repulsive forces prevent it from approaching obstacles. The model and the obstacles act together as a positive charge, while the target position acts as a negative charge. The repulsive effect between the forces of the model and those of the obstacles increases as they get closer to each other and decreases as they recede, thereby guiding the robot along a safe path to its goal. The effects of the attractive and repulsive potentials are illustrated in Figure 2. Figure 2a illustrates the potentials in the field without considering the repulsive potentials
from the obstacles, while the combination of potentials is presented in Figure 2b.

\begin{figure*}[!ht]
\centering
\includegraphics[width=0.65\textwidth]{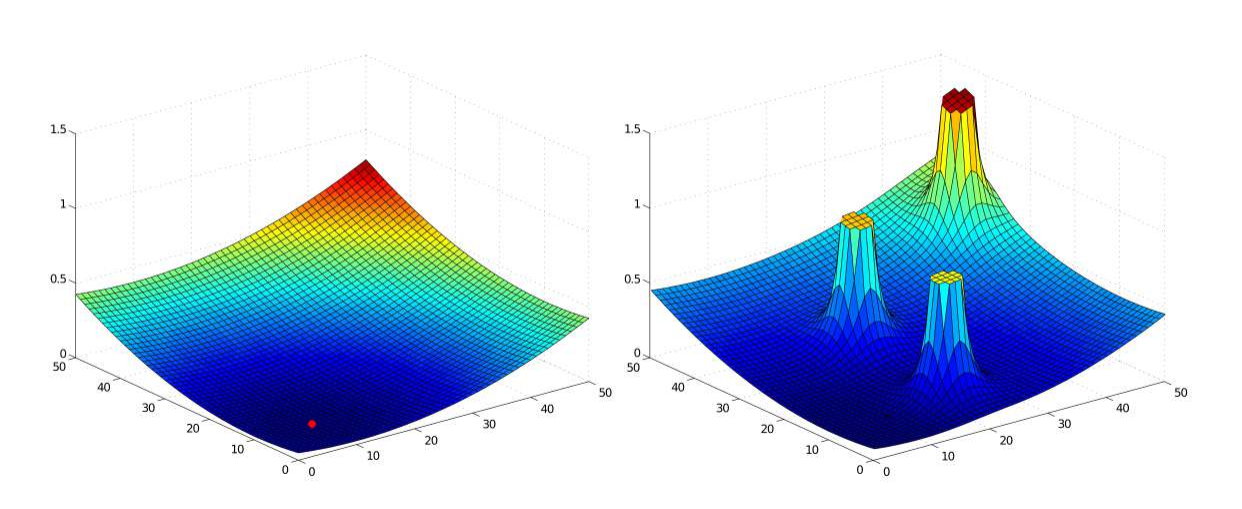}
\caption[Visual Overview of Potentials: a) Attractive Potential b) Combination of the Attractive and the Repulsive Potential]{Visual Overview of Potentials: a) Attractive b) Combination of the Attractive and the Repulsive Potentials \cite{Obstacle2005Ribeiro}}
\label{PotentialFieldsCombined}
\end{figure*}

The implementation for realizing obstacle avoidance involves a subscriber to the topic ’/spur/laser/scan’, which evaluates the data collected by the rays. Upon updating the acquisition, the logic performs verification of whether the defined tolerance range for the distance has been exceeded. In such an event, the specified potentials are applied and an avoidance vector is generated. The resulting target position is used to command the model to travel to a safe
position in its environment.

\subsection{Automatic Waypoint Navigation}

In the Gazebo simulator, the internal coordinates [X, Y, Z, \begin{math}\psi\end{math} ] are assigned to the respective position within the local frame, which is established for each model based on its location prior to the initial takeoff.
As a consequence, no notion of geo-referenced coordinates is realized in the Gazebo simulator. In order to implement an autonomous simulation, it is necessary to instruct the model to move along its internal frame by a specific value. The attainment of a target position can be detected by continuously verifying the location of the
drone and comparing it to the target position with respect to the drone’s position prior to the execution of the instruction. The function \textit{set\_destination} uses the X, Y, Z coordinates, and the rotation parameter related to the position at the specific time of the instruction execution in order to control the drone.

An excerpt of the most significant implementation steps of the function is provided in Listing \ref{lst:setDestination}. The function compiles the destination coordinates on the basis of three components, while the odometry message of the \textit{nav\_msgs} topic provides information about the estimated position and velocity. By utilizing the resulting \textit{heading} of the model (line 155), the coordinates are transformed from map coordinates to local coordinates (lines 156 - 157). This allows the current model location, with valid local coordinates, to be supplemented with the current position on the map.

\input{Listings/setDestination.tex}

In order to assemble the obtained coordinates to create a waypoint, they need to be compiled into a variable of the type PoseStamped\footnote{
\url{http://docs.ros.org/en/api/geometry_msgs/html/msg/PoseStamped.html} [Accessed May-2022]} message, which is part of the \textit{geometry\_msgs} package. Finally, the orchestrated waypoint consisting of the calculated X, Y, and Z coordinates is published to the topic \textit{local\_pos\_pub}, according to which the drone acts as a subscriber and thus receives the instruction.

In order to determine whether the target position has been reached by the model, the function \textit{check\_waypoint\_reached} can be utilized, wherein the parameters for the specification of tolerance values for both the position and the heading of the model can be specified. To determine the current deviation of the position data from the target position, the absolute differences of the target
coordinates to the current position data are generated. The delta between the two points is calculated by a single application of the Pythagorean theorem. In higher dimensions, the Euclidean distance of n Cartesian coordinates can be obtained by the multiple application of the Pythagorean theorem \cite{Geometry2004Tabak}. An example of how the theorem can be used for three-dimensional space is presented in Listing \ref{lst:checkWaypointReached} line 352.

% \begin{gather}
% \label{eq-1}
% 	distance = \sqrt{(\Delta X)^{2} + (\Delta Y)^{2} + (\Delta Z)^{2}} 
% \end{gather}

The aforementioned functions are used by a control program for the simulation of a waypoint logic, which is provided by the GNC Library. The control program uses the functions for initializing the ROS-main node as well as the publisher and subscriber method. After the Flight Control Unit (FCU) has successfully established a connection and the takeoff was enabled by setting the model into the GUIDED mode, the local
reference frame is built up until the drone receives the instruction to takeoff. The logic contains a vector list of waypoints, which are composed of the information for
X-, Y-, Z-axis, and \begin{math}\psi\end{math} for each position. As long as the main node of ROS is active, the drone is instructed by the \textit{set\_destination} function to head for the next target in the list. Based on the ROS rate,
the \textit{check\_waypoint\_reached} function periodically checks whether the target
position has been reached (see Listing \ref{lst:checkWaypointReached}). 

In order to facilitate that the attack on the odometry leads to the intended result, changes had to be made regarding the functions of the API. The function \textit{set\_destination} provides the target position with three
values, which are given below using the example of the X-coordinate:

\begin{enumerate}
   \item \textit{'local\_offset\_pose\_g.x'}: Specifies the current position in the local reference frame. The value is read by the API from the topic '\textit{/mavros/global\_position/local}'. The manipulation by an attacker using a script is conducted in a way that the original and correct value is no longer read, but instead, the spoofed value is obtained for that variable. 
   \item \textit{'Xlocal'}: Contains the value of the waypoint vector, which has been converted from map coordinates to local coordinates. This was done in order for it to be used in the simulation. 
   \item \textit{'correction\_vector\_g.position.x'}: Contains a correction value for compensation of position mapping errors. The experiments conducted as a part of this research applied $$\lim_{\text{cv}\to 0}$$ with $cv$ representing the $correction\_vector\_g.position$. 
\end{enumerate}

\input{Listings/checkWaypointReached}

By default, the API performs the addition of these values. Consequently, a point specified in the waypoint vector is not directly approached, but rather a target
coordinate is constructed from the specified point as well as the current location. If
the model’s position is not displaced by the origin and is not influenced by an external
force, this form of implementation is functional. However, in order to observe the impact of the attacks on the odometry of the model in the simulation in a form
that would occur in the event of a physical attack in a real setting, the computation of the target position had to be adjusted. To determine the coordinates of the target position, a subtotal of the current local position and the corresponding correction value is obtained. In the next step, the delta of this value with the point that is to be approached is determined. Finally, the differential value of each coordinate is added to the current position of the model.

After everything has been initialized, the local frame is started and the takeoff of the
model is performed. At this point, the attack on the odometry may be performed, whereby the manipulated position data is adopted for further execution.

%% file: Listings/setDestination.tex
\begin{lstlisting}[label=lst:setDestination, 
caption = Excerpt of the function \textit{set\_destination}, 
firstnumber = 18, 
breakindent = 50pt, 
language = Python,
xleftmargin = 1.3em,
linewidth = 8.4cm]
geometry_msgs::PoseStamped waypoint_g; (*@ \stopnumber @*)
(*@ \startnumber{41} @*) 
ros::Publisher local_pos_pub; (*@ \stopnumber @*)
... (*@ \startnumber{150} @*) 
void set_destination(float x, float y, float z, float psi)
{ (*@ \stopnumber @*)
    ... (*@ \startnumber{154} @*) 
    float deg2rad = (M_PI/180);
    float heading = (correction_heading_g + local_offset_g - 90)*deg2rad;
    float Xlocal = x*cos(heading) - y*sin(heading);
    float Ylocal = x*sin(heading) + y*cos(heading);(*@ \stopnumber @*)
    ... (*@ \startnumber{159} @*)
    x = (difference(Xlocal, (local_offset_pose_g.x + correction_vector_g.position.x))) +           correction_vector_g.position.x;
    y = (difference(Ylocal, (local_offset_pose_g.y + correction_vector_g.position.y))) +           correction_vector_g.position.y;
    z = (difference(Zlocal, (local_offset_pose_g.z + correction_vector_g.position.z))) +           correction_vector_g.position.z; (*@ \stopnumber @*)
    ... (*@ \startnumber{164} @*)
    waypoint_g.pose.position.x = x;
    waypoint_g.pose.position.y = y;
    waypoint_g.pose.position.z = z;
        
    local_pos_pub.publish(waypoint_g);
    
\end{lstlisting}

%% file: Listings/checkWaypointReached.tex
\begin{lstlisting}[label=lst:checkWaypointReached, caption = Excerpt of the function \textit{check\_waypoint\_reached} from the GNC library, captionpos=b, firstnumber = 18, 
breakindent = 50pt, 
language = Python,
xleftmargin = 1.3em,
linewidth = 8.4cm]
int check_waypoint_reached(float pos_tolerance=0.3, float heading_tolerance=0.01)
{ (*@ \stopnumber @*) (*@ \startnumber{349} @*) 
    float deltaX = abs(waypoint_g.pose.position.x -           current_pose_g.pose.pose.position.x);
    float deltaY = abs(waypoint_g.pose.position.y -           current_pose_g.pose.pose.position.y);
    float deltaZ = abs(waypoint_g.pose.position.z -           current_pose_g.pose.pose.position.z);
    float dMag = sqrt(pow(deltaX, 2) + pow(deltaY, 2) + pow(deltaZ, 2)); (*@ \stopnumber @*)
    ... (*@ \startnumber{357} @*)    
    float cosErr = cos(current_heading_g*deg2rad) -                      cos(local_desired_heading_g*deg2rad);
    float sinErr = sin(current_heading_g*deg2rad) -                      sin(local_desired_heading_g*deg2rad); 
    
    float headingErr = sqrt( pow(cosErr, 2) + pow(sinErr, 2) ); (*@ \stopnumber @*)
    ... (*@ \startnumber{366} @*) 
    if(dMag < pos_tolerance && headingErr < heading_tolerance)
    {
        return 1; (*@ \stopnumber @*)
\end{lstlisting}

%% file: Chapters/AttackSimulations.tex
\newcommand{\topiclidar}{\textit{/spur/laser/scan}}
\newcommand{\topicnavloc}{\textit{/mavros/global\_position/local}}
\newcommand{\topicnavglob}{\textit{/mavros/global\_position/global}} 

This section presents an approach to simulate spoofing and jamming attacks that target sensors. The simulated attacks include scenarios in which an attacker targets the GPS component as well as the LiDAR system. The approach is evaluated by simulating the aforementioned attacks using the GNC API (see Section \ref{sec:gnc}). 

In order to simulate the LiDAR component, an obstacle avoidance algorithm is utilized as part of a simulation provided by the API (see Section \ref{sec:exp}). This means that the model is able to determine the closest distance to obstacles and avoid them autonomously. Consequently, the developed attacks
are tested against the aforementioned waypoint program, which enables the execution of an autonomous flight maneuver.
They are implemented in the form of independently executable scripts that allow them to be executed independently of their field of application. The values to be spoofed or jammed are inquired via the terminal during the execution of the script. In the following, the implemented attack scenarios that are part of this research are presented in more detail: 

\vspace{\baselineskip} %\noindent
%\textbf{LiDAR}:
LiDAR:
Regardless of the specific attack scenario, the objective of the attacks targeting the LiDAR component is to render the model incapable of correctly detecting an object within the sensor's range. As a result, the sensor and subsequently the obstacle avoidance logic are no longer able to detect objects correctly. This is described in the following three scenarios (Scenario 1-3). \textbf{Scenario 1} causes the distance values of the LiDAR sensor to be spoofed with values outside the range of the obstacle avoidance algorithm. Hence, the system is no longer able to detect any obstacles in its environment. In the \textbf{Scenario 2}, the distance values are spoofed with other values that are within the threshold of the avoidance algorithm. This results in unpredictable behavior originating from the implementation and mathematical background of the avoidance algorithm. Likewise to the first attack scenario, the reliable detection of objects is no longer possible. \textbf{Scenario 3} demonstrates the simulation of a more sophisticated attack. Instead of spoofing all the distance values of the LiDAR's range using uniform values, a specific range is selectively targeted. Within this range, the distance values are spoofed using values within the obstacle avoidance threshold. The values of the remaining detection range are spoofed with values outside the threshold. This attack is intended to model the existence of an object in the model's environment. When the avoidance algorithm is active, this scenario allows an attacker to purposely control the model.

\vspace{\baselineskip} %\noindent
%\textbf{GPS}:
GPS:
The development of the attacks that target the odometry-related data is closely related to the ones targeting the LiDAR’s data. For such an attack to be successful, the position data needs to be targeted by an attacker. This allows for the manipulations of the current location of the model by falsifying the latitude and longitude data of the GPS component's measurements. This means that the internal position data is updated with the manipulated data. The aforementioned internal position data of ROS is then used for the coordination of models in the simulation. Hence, this makes it possible to observe the effects of an attack implemented as a simulation. However, in case the data of the GPS component  was solely manipulated, it would not be possible to observe the effects of the attacks in a simulation without additional developments. 
% make sure that the internal position data is explained in the background section
In the course of this research, two main attack scenarios (Scenario 1 and Scenario 2) that target the GPS component were implemented. \textbf{Scenario 1:} The arbitrary data is spoofed regarding the location of the model. As a result, it is no longer feasible for ROS nodes to obtain the authentic position. The waypoint logic of the GNC API is therefore no longer able to verify whether a waypoint has been reached. \textbf{Scenario 2:} The odometry is deliberately manipulated as if the drone was located at a specific location that is displaced from its actual position. The execution of this attack results in the drone heading in the opposite direction of the waypoint that should be targeted.

% shows the effect of the manipulation of the local position
%\begin{figure}%[h]
%    \begin{center}
 %      \includegraphics[width=\linewidth]{Figures/OdometryAttackII.eps}
%    \end{center}
%    \caption[Visualisation of the Effects of Manipulating the Local Position by a Specific Offset]{Visualisation of the Effects of Manipulating the Local Position by a Specific Offset}
%    \label{fig:obstacleAvoidanceInfThreshold}
%\end{figure}

\begin{figure}%[h]
    \begin{center}
        \includegraphics[width=0.8\linewidth]{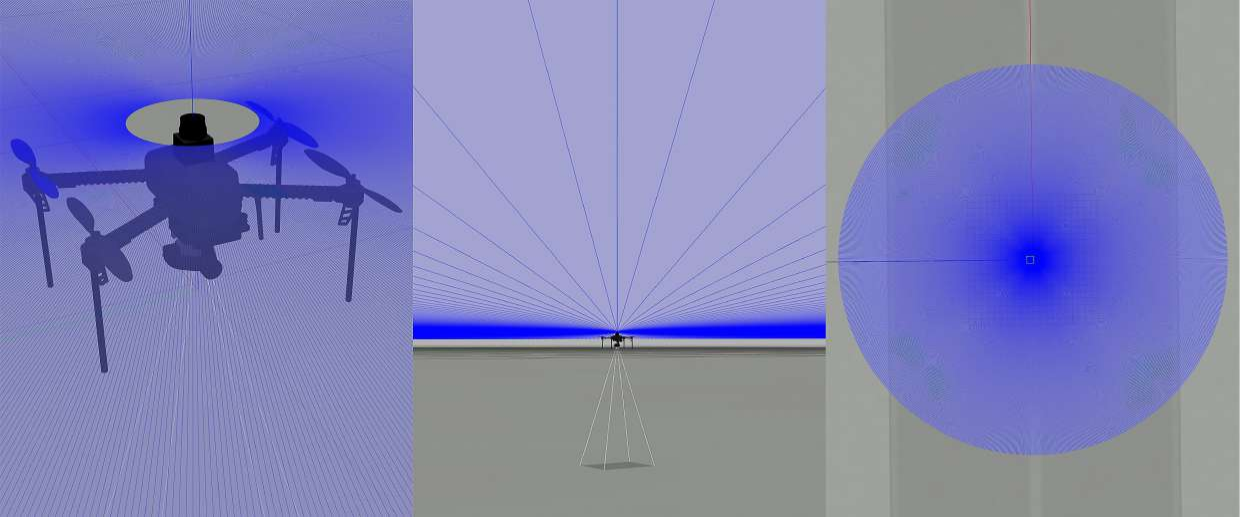}
    \end{center}
    \caption[360° LiDAR Sensor on the Top of the Iris Model, Frontal View of the Rays and Top-down View of the Rays' Range (Left to Right)]{360° LiDAR Sensor on the Top of the Iris Model, Frontal View of the Rays and Top-down View of the Rays' Range (Left to Right)}
    \label{fig:rayDistribution}
\end{figure}

\input{Listings/LidarSpoofing.tex}

In the case of a jamming attack in the physical world, noise signals with a signal power higher than that of the original signal are transmitted to overpower it. As a consequence, these counterfeit signals are interpreted as valid and their data is passed on for further processing. 
To simulate this phenomenon in the proposed approach, the data retrieved by the sensor in the implementation is reproduced. This was done by mimicking the data structure of the genuine signal information and publishing it at an increased rate on the corresponding ROS topic. 
The circumstances of the generation of a simple jamming signal targeting the LiDAR topic are given in Listing \ref{lst:lidarSpoofing}. The LiDAR sensor used, and therefore also the authentic signal, covers 1024 rays in its full detection range. The distribution of the rays in the case of a 360° LiDAR is shown in Figure \ref{fig:rayDistribution}. In simulations, the LiDAR plugin \textit{hokuyo\_node}\footnote{Further details about the plugin:
\url{http://wiki.ros.org/hokuyo\_node} [Accessed Aug-2022]} was added to the model. The plugin defines the x-axis along the zero-degree ray and the angle direction as clockwise positive by default. 

The distances that are acquired by the rays are mapped and stored in a tuple. The counterfeit tuple contains the values passed by an attacker when executing an attack. In case of a jamming attack, the tuple is filled with random values within a specified range. By utilizing these incidental values, the characteristic of a noise signal is supposed to be imitated. 
The genuine data is published to the topic via ROS messages at the rate defined in the configuration of the sensor. Derived from previous research by 
Kerns et al. \cite{kerns2014unmanned} and 
Rozenbeek et al. \cite{rozenbeek2020evaluation} 
the power of the counterfeit signal must be sufficiently higher than the genuine signal. 
% I would very much like to give exact data and numbers here, however no literature I have read comments on this with specific numbers. 
By exploiting this concept, the implemented logic adopts a ROS rate that is a hundred times higher than the authentic rates of the sensor's data output. The
majority of sensors currently offered by the manufacturer Hokuyo require 100ms for scanning their full detection range. To perform a full update of the data, this would result in a frequency of 10Hz.
% Ensure that ROS is written about rates in the background.
During the development, different rates were tested experimentally. The effects of the attacks could already be observed at low multiples of the rate of the genuine signal. At the selected frequency of 1000Hz, it was possible to observe the effects in the simulations that could create reliability issues.

The first two attacks against the LiDAR sensor can already be realized utilizing the previously described implementation. The distance values stored by the tuples are set to values either within or outside the range of the obstacle avoidance threshold. The generated data is then published to the ROS topic '\topiclidar' 
% The ' ' are necessary so that afterwards also again a space is made
at a high rate as previously described. 

\begin{figure}%[h]
    \begin{center}
       \includegraphics[width=0.5\linewidth]{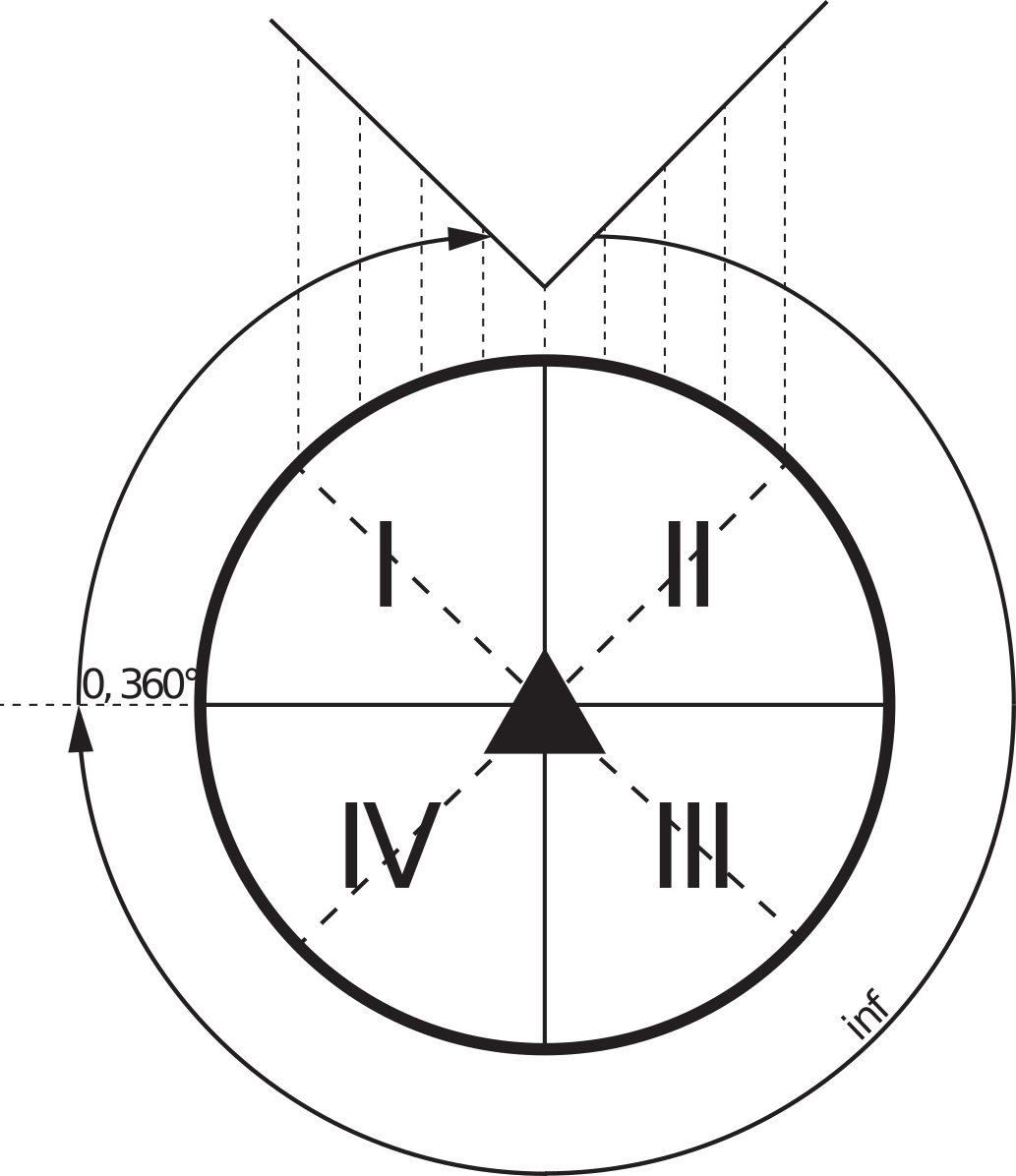}
       % Alternatively: OdometryAttackObstalceWithoutQuarters.eps
       % same Figure without the Numbering of the Quarters
    \end{center}
    \caption[Visualization of the rays manipulated with distance values within the threshold to model a non-existent front-facing object. ]{Visualization of the rays manipulated with distance values within the threshold to model a non-existent front-facing object. }
    \label{fig:odometryAttackObstacle}
\end{figure}

For the implementation of the more sophisticated attack scenario, in which a non-existent object is modeled, the content of the tuple needs to be generated selectively. Therefore, the rays that are most forward-oriented with respect to the model must be assigned with distance values within the avoidance threshold.
%Given the configuration of the sensor, this corresponds to the last reading of the second quarter's second half and the first reading at the beginning of the third quarter's first half. %the previous sentence is not clear at all
The applied LiDAR sensor utilizes 1024 rays to cover a range of 360°. For the implementation of attacks, the rays were divided into eighths alongside the circuit of 128 rays each. This is presented in Figure \ref{fig:odometryAttackObstacle}. The two rays at the frontmost tip of the model are manipulated with a distance value given by an attacker. These are the last ray of the second eighth and the first ray of the third eighth of the coverage area. 
This is achieved by applying linear gradation between these rays and the boundaries of their eighths. By manipulating the range tuples in this manner, the obstacle avoidance logic is expected to be triggered. As a result, the model moves backward, which is the direct cause of the executed attack. 

\input{Listings/gpsSpoofing.tex}

For the injection of counterfeit messages that target the odometry of a ROS model, the following should be taken into consideration: 
Many models in ROS, including the Iris model, are equipped with a GPS component by default. The component is represented by the message \textit{sensor\_msgs} used for the mapping of global coordinates. The message is published on its corresponding topic 
'\topicnavglob'\footnote{\url{http://docs.ros.org/en/lunar/api/sensor_msgs/html/msg/NavSatFix.html} [Accessed Aug-2022]}. The coordinates are determined by the message using two fields for both latitude and longitude. 
For internal navigation instructions in ROS, another message of the type anv\_msgs\footnote{\url{http://docs.ros.org/en/noetic/api/nav_msgs/html/msg/Odometry.html} [Accessed Aug-2022]} is used. It refers to an odometry message that indicates the position with respect to the local frame and is published to the topic '\topicnavloc'. This is required as each model in ROS has internal coordinates assigned to it, which are used for coordination.
For representing a physical spoofing attack targeting the GPS component, it is probably necessary to manipulate the values of the topic '\topicnavglob' with respective values for the longitude and latitude. Following the scheme of the attacks on the LiDAR component, a message containing values for the longitude and latitude is generated and published at a higher rate compared to the accordant topic. The spoofing of the messages of the topic '\topicnavglob' is achieved by an adversary transmitting longitude and latitude values. Since the values of the internal coordinates are further used for the coordination, the values of the NavSatFix\footnote{\url{http://docs.ros.org/en/melodic/api/sensor_msgs/html/msg/NavSatFix.html} [Accessed Aug-2022]} message have to be propagated to the odometry message and published to the '\topicnavloc' topic. Consequently, the aforementioned scenario is demonstrated in Listing \ref{lst:gpsSpoofing}.

The distinction made in the implementation between spoofing and jamming analogously follows the approach taken for the implementation of the LiDAR attacks. The jamming is implemented by assigning random values within a range defined by an attacker. 
By deliberately manipulating the messages of the '\topicnavglob' and '\topicnavloc' topics, the actual position data can no longer be correctly evaluated. Since it is no longer possible to obtain the actual position data, the model is no longer able to verify the attainment of a given waypoint. 

%% file: Listings/LidarSpoofing.tex
\begin{minipage}{\linewidth}
% Excerpt of the script %\textbf{lidar_spoofing.py} with the implementation for the manipulation of the LiDAR topic ('\topiclidar')
\begin{lstlisting}[label=lst:lidarSpoofing, 
caption = Tuple Generation Using Arbitrary Values, 
firstnumber = 18, 
breakindent = 50pt, 
language = Python,
xleftmargin = 0.8em,
linewidth = 8.4cm
]
rate = rospy.Rate(1000) (*@ \stopnumber @*)
(*@ \startnumber{21} @*) 
for x in range(1024):
    y.append(float(distance))
thistuple = tuple(y)
laser.ranges = thistuple (*@ \stopnumber @*)
(*@ \startnumber{25} @*) 
while not rospy.is_shutdown():
    scan_pub.publish(laser)
    rate.sleep()
\end{lstlisting} 
\end{minipage}

%% file: Listings/gpsSpoofing.tex
\begin{lstlisting}[label=lst:gpsSpoofing, 
caption = Excerpt of the script \textit{gps\_spoofing.py} showing the implementation for the manipulation of the odometry topics,
firstnumber = 18, 
breakindent = 50pt, 
language = Python,
xleftmargin = 1.3em,
linewidth = 8.4cm]
global_publisher = rospy.Publisher('/mavros/global_position/global', NavSatFix, queue_size=10)
local_publisher = rospy.Publisher('/mavros/global_position/local', Odometry, queue_size=10) (*@ \stopnumber @*)
(*@ \startnumber{20} @*) 
global_message.longitude = float(longitude)
global_message.latitude = float(latitude) (*@ \stopnumber @*)
(*@ \startnumber{24} @*) 
local_message.pose.pose.position.x = global_message.longitude
local_message.pose.pose.position.y = global_message.latitude (*@ \stopnumber @*)
(*@ \startnumber{30} @*) 
def spoof_local():
    local_publisher.publish(local_message) (*@ \stopnumber @*)
    (*@ \startnumber{37} @*) 
def spoof_global():
    global_publisher.publish(global_message) (*@ \stopnumber @*)
\end{lstlisting}

%% file: Chapters/Evaluation.tex
In order to evaluate the proposed approach, the scenarios developed in Section \ref{sec:attsim} were tested against the simulations provided by the GNC Library. In the following section, we describe the behavior that was observed during the execution of each attack scenario.

\textbf{LiDAR:} In the first scenario when the messages with distance values outside the threshold were injected, an unpredictable behavior was observed. This behavior results from the generated avoidance vectors of the avoidance algorithm. In the second scenario, the messages with distance values within the threshold were injected into the LiDAR topic. Without executing an attack, the model avoids objects that are close to the model during the simulation when the obstacle avoidance logic is activated. As soon as the attack is executed, the generated messages are published to the LiDAR topic. 
When executing an attack where distance values outside the avoidance threshold were specified, it was possible to observe that the obstacle avoidance logic was no longer triggered. If an object is subsequently introduced into the scene that lies within the threshold, the model would no longer avoid it. Due to the injected messages with counterfeit data, the sensor's correct data can no longer be evaluated. If the model is instructed to move to a position where another object is located, a crash would occur. Finally, the execution of the attack, in which certain rays were specifically assigned with distance values within the threshold, was checked against the waypoint logic. It was expected that the model could be steered in a certain direction by the attack. The experiments showed that this behavior occurred at the beginning of the attack. The model steered away to the rear as if an object was close to its front. With the distance increasing, it was possible to see a deviation in the model's flight path.

\textbf{GPS:} In the first scenario, messages with arbitrary values were injected on both the topic storing the global and the topic containing the local position data. In the course of autonomous waypoint navigation, reaching a target waypoint is determined through the comparison of the current position of the model with the target position. By executing the attack, this comparison is no longer possible. This is because the evaluation of the attack started right after the launch of the waypoint logic. As a result, it was possible to observe that the model approached the position of the first waypoint. Upon arrival, no further maneuvers were performed by the model, and no landing occurred. Thus, the model remained hovered in the air. In the second scenario, the location data was manipulated as if the model was at the position displaced in a certain direction from its actual location. As a result, an attacker is able to take control over the model to some extent. In one of the test cases that are a part of this evaluation, the position data was manipulated in such a way that it was initially shifted by the value 10 on the X-axis from its own position. Subsequently, the model was instructed by the waypoint logic to move to a point that was offset along the X-axis by the positive value 5. We observed that the model no longer moved along the X-axis with increasing values, but rather traveled in the opposite direction.

%% file: Chapters/Discussion.tex
In this section, we discuss the key findings of the implemented attack simulations as well as the difference between conducting simulations against real-world evaluation.

\subsection{Key Findings}
In the following subsection, key findings related to conducted simulations of attacks on both LiDAR and GPS components are discussed.

\textbf{LiDAR:} Spoofing or jamming distance values beyond the obstacle avoidance threshold results in a scenario, wherein objects placed within the threshold, no longer trigger the avoidance mechanism. The reason behind this particular behavior can be explained as
follows: The obstacle avoidance logic’s topic is no longer able to retrieve correct values, as it is spoofed by the node of the attack script that contains incorrect values. When spoofing or jamming distance values within the obstacle avoidance threshold, the drone moves in a circular motion around its original point. Both of the addressed simulations, wherein possibilities for an attacker to
indirectly control a vehicle, led to a result in which the flight path could potentially be manipulated. Consequently, it was determined that the behavior
of the UAV cannot be predicted in advance. Because of this, a clear effect of the attack can be observed in this case as well. On this occasion, it should be pointed
out that in cases when the scenarios are carried out after a given distance, a deviation of the trajectory in a circular path towards the origin could be observed. The deviation results are the same as described by Ribeiro \cite{Obstacle2005Ribeiro} from the properties of the field
method presented by Khatib \cite{Realtime1995Khatib}, which is used to determine the avoidance
vectors. The exact theoretical reasons that cause this behavior would require a more detailed analysis of the algorithm and other possible alternatives. However, since the analysis of obstacle avoidance algorithms is not the main focus of this research, it was not investigated in more detail. 

\textbf{GPS:} Spoofing or jamming the odometric data does not change the fact that the first set waypoint is still reached. However, once the UAV is in the position of the waypoint, no further maneuvers are performed by the model and landing does not occur. The reason behind this resulting behavior after an attack has been executed is the topic related to the waypoint logic. It retrieves the data regarding the current position from the topic with the local position data. However, due to the fact that the attack script specifically spoofs these topics with incorrect values, it is no longer possible to determine the current valid position. As a result, the GNC API cannot verify whether a targeted waypoint has been reached. Thus, all the other waypoints have never been approached, and landing has not been initiated. It might explain why, in the first scenario, the model acted as described previously. Due to the adjustments of the control logic outlined in Section \ref{sec:attsim}, another malicious node could be introduced, which allows the UAV to be controlled by an adversary to a certain extent. If an attacker is able to do this and spoof an opposite position, the flight path of the model could be altered indirectly. As long as the topic containing the local position data is spoofed with malicious values, it also becomes unfeasible for the position of a drone to be verified in this scenario. Similar to the first scenario, no further waypoints are approached in this specific situation as well.

\subsection{Simulation vs Real-world Evaluation}

The preceding demonstration of the implemented attack scenarios clearly demonstrates that it is feasible to simulate jamming and spoofing attacks on a ROS-developed system
using the Gazebo simulator. In order to verify the feasibility of this research, we compared our results to the results obtained by Pokhrel \cite{Drone2018Pokhrel}, who proposed an approach for drone obstacle avoidance and navigation using artificial intelligence. Similar to our approach, Pokhrel \cite{Drone2018Pokhrel}
also conducted simulations using the ROS and Gazebo simulator as part of his investigations. In addition, he created a physical setup to experimentally evaluate his investigations. This allowed a comparison between the results of the simulation and the outcome of the physical real-world experiment. The results showed that there was a difference in the noise perception of the sensors between the two setups. The noise in the real world was found to be higher than the one recorded in the simulations performed in the Gazebo simulator. Thus, it can be deduced that the Signal to Noise Ratio (SNR) has a certain impact when investigating the effects of jamming and spoofing attacks on LiDAR and GPS components. However, it is still beneficial to initially conduct simulations of such attacks in a simulator to avoid any major safety impacts. Nevertheless, to completely and critically evaluate the complete benefits of the presented simulations, it is necessary to perform a full experimental evaluation of the
scenarios in the physical world.

%% file: Chapters/Conclusion.tex
This paper presented a novel approach to simulate attacks that target sensor-related components of UAVs. The addressed attacks include jamming and spoofing attacks that are aimed to affect GPS and LiDAR components. As a starting point, we investigated both the ROS framework and the MAVLink protocol, which were analyzed regarding potential vulnerabilities. In order to conduct attack simulations, the vulnerability that relates to the processes’ data exchange in ROS applications was exploited. According to the analyzed research that is part of the related work (see Section \ref{sec:rw}), several attack scenarios were identified that were simulated in the Gazebo simulator. This was achieved by implementing malicious processes, which specifically manipulate the data of the corresponding topics. 

Despite all the benefits that attack simulations bring, such as testing in a safe environment, it is still advisable to conduct real-world testing. Consequently, this makes it possible to compare the simulation results against the results obtained from real-world experiments. The future work includes developing more sophisticated attack simulations. In addition to generating attacks that target components such as GPS and LiDAR, it is necessary to develop other simulations that would also affect other components such as radar, inertial measurement unit, and models that perform analysis of a camera’s image.

%% file: main.bbl
% Generated by IEEEtran.bst, version: 1.14 (2015/08/26)
\begin{thebibliography}{10}
\providecommand{\url}[1]{#1}
\csname url@samestyle\endcsname
\providecommand{\newblock}{\relax}
\providecommand{\bibinfo}[2]{#2}
\providecommand{\BIBentrySTDinterwordspacing}{\spaceskip=0pt\relax}
\providecommand{\BIBentryALTinterwordstretchfactor}{4}
\providecommand{\BIBentryALTinterwordspacing}{\spaceskip=\fontdimen2\font plus
\BIBentryALTinterwordstretchfactor\fontdimen3\font minus
  \fontdimen4\font\relax}
\providecommand{\BIBforeignlanguage}[2]{{%
\expandafter\ifx\csname l@#1\endcsname\relax
\typeout{** WARNING: IEEEtran.bst: No hyphenation pattern has been}%
\typeout{** loaded for the language `#1'. Using the pattern for}%
\typeout{** the default language instead.}%
\else
\language=\csname l@#1\endcsname
\fi
#2}}
\providecommand{\BIBdecl}{\relax}
\BIBdecl

\bibitem{doi/10.2829/085259}
\emph{Single European Sky ATM Research 3 Joint Undertaking. European drones
  outlook study : unlocking the value for Europe}.\hskip 1em plus 0.5em minus
  0.4em\relax Publications Office, 2017.

\bibitem{ramos2021security}
S.~Ramos, T.~Cruz, and P.~Sim{\~o}es, ``Security and safety of unmanned air
  vehicles: An overview,'' in \emph{ECCWS 2021 20th European Conference on
  Cyber Warfare and Security}.\hskip 1em plus 0.5em minus 0.4em\relax Academic
  Conferences Inter Ltd, 2021, p. 357.

\bibitem{veerappan2021drat}
C.~S. Veerappan, P.~L.~K. Keong, V.~Balachandran, and M.~S. B.~M. Fadilah,
  ``Drat: A penetration testing framework for drones,'' in \emph{2021 IEEE 16th
  Conference on Industrial Electronics and Applications (ICIEA)}.\hskip 1em
  plus 0.5em minus 0.4em\relax IEEE, 2021, pp. 498--503.

\bibitem{gazebo}
``{Gazebo Simulator},'' \url{https://gazebosim.org/home}, accessed: 2022-08-20.

\bibitem{witte2022towards}
T.~Witte, R.~Groner, A.~Raschke, M.~Tichy, I.~Pekaric, and M.~Felderer,
  ``Towards model co-evolution across self-adaptation steps for combined safety
  and security analysis,'' EasyChair, Tech. Rep., 2022.

\bibitem{qian2014manipulation}
W.~Qian, Z.~Xia, J.~Xiong, Y.~Gan, Y.~Guo, S.~Weng, H.~Deng, Y.~Hu, and
  J.~Zhang, ``Manipulation task simulation using ros and gazebo,'' in
  \emph{2014 IEEE International Conference on Robotics and Biomimetics (ROBIO
  2014)}.\hskip 1em plus 0.5em minus 0.4em\relax IEEE, 2014, pp. 2594--2598.

\bibitem{takaya2016simulation}
K.~Takaya, T.~Asai, V.~Kroumov, and F.~Smarandache, ``Simulation environment
  for mobile robots testing using ros and gazebo,'' in \emph{2016 20th
  International Conference on System Theory, Control and Computing
  (ICSTCC)}.\hskip 1em plus 0.5em minus 0.4em\relax IEEE, 2016, pp. 96--101.

\bibitem{o2014gentle}
J.~M. O'Kane, ``A gentle introduction to ros,'' 2014.

\bibitem{raje2020evaluation}
S.~Raje, ``Evaluation of ros and gazebo simulation environment using turtlebot3
  robot,'' 2020.

\bibitem{reutebuch2005light}
S.~E. Reutebuch, H.-E. Andersen, and R.~J. McGaughey, ``Light detection and
  ranging (lidar): an emerging tool for multiple resource inventory,''
  \emph{Journal of forestry}, vol. 103, no.~6, pp. 286--292, 2005.

\bibitem{khan2021comparative}
M.~U. Khan, S.~A.~A. Zaidi, A.~Ishtiaq, S.~U.~R. Bukhari, S.~Samer, and
  A.~Farman, ``A comparative survey of lidar-slam and lidar based sensor
  technologies,'' in \emph{2021 Mohammad Ali Jinnah University International
  Conference on Computing (MAJICC)}.\hskip 1em plus 0.5em minus 0.4em\relax
  IEEE, 2021, pp. 1--8.

\bibitem{wells1987guide}
D.~Wells, N.~Beck, A.~Kleusberg, E.~J. Krakiwsky, G.~Lachapelle, R.~B. Langley,
  K.-p. Schwarz, J.~M. Tranquilla, P.~Vanicek, and D.~Delikaraoglou, ``Guide to
  gps positioning,'' in \emph{Canadian GPS Assoc}.\hskip 1em plus 0.5em minus
  0.4em\relax Citeseer, 1987.

\bibitem{tiesinga2021codata}
E.~Tiesinga, P.~J. Mohr, D.~B. Newell, and B.~N. Taylor, ``Codata recommended
  values of the fundamental physical constants: 2018,'' \emph{Journal of
  Physical and Chemical Reference Data}, vol.~50, no.~3, p. 033105, 2021.

\bibitem{Evaluation2020Rozenbeek}
D.~J. Rozenbeek, ``Evaluation of drone neutralization methods using radio
  jamming and spoofing techniques,'' {PhD} dissertation, School of Electrical
  Engineering and Computer Science. Stockholm, Sweden, 2020.

\bibitem{IntelligentGnc2022Johnson}
E.~Johnson, S.~Annath, M.~Burri, and M.~Kamel, ``Intelligent quads drone
  development software collection,''
  \url{https://github.com/Intelligent-Quads/iq_gnc}, 2019, online: Github
  repository [Accessed: February-2022]".

\bibitem{IntelligentSim2022Johnson}
E.~Johnson, S.~Annath, F.~Furrer, and M.~Kamel, ``Intelligent quads
  simulations,'' \url{https://github.com/Intelligent-Quads/iq_sim}, 2018,
  online: Github repository [Accessed: February-2022]".

\bibitem{Programming2014Bartonicek}
J.~Bartoníček, ``Programming language paradigms \& the main principles of
  object-oriented programming,'' \emph{CRIS - Bulletin of the Centre for
  Research and Interdisciplinary Study}, vol. 2014, pp. 1--7, 2014.

\bibitem{kerns2014unmanned}
A.~J. Kerns, D.~P. Shepard, J.~A. Bhatti, and T.~E. Humphreys, ``Unmanned
  aircraft capture and control via gps spoofing,'' \emph{Journal of Field
  Robotics}, vol.~31, no.~4, pp. 617--636, 2014.

\bibitem{eldosouky2019drones}
A.~Eldosouky, A.~Ferdowsi, and W.~Saad, ``Drones in distress: A game-theoretic
  countermeasure for protecting uavs against gps spoofing,'' \emph{IEEE
  Internet of Things Journal}, vol.~7, no.~4, pp. 2840--2854, 2019.

\bibitem{gaspar2020capture}
J.~Gaspar, R.~Ferreira, P.~Sebasti{\~a}o, and N.~Souto, ``Capture of uavs
  through gps spoofing using low-cost sdr platforms,'' \emph{Wireless Personal
  Communications}, vol. 115, no.~4, pp. 2729--2754, 2020.

\bibitem{petit2015remote}
J.~Petit, B.~Stottelaar, M.~Feiri, and F.~Kargl, ``Remote attacks on automated
  vehicles sensors: Experiments on camera and lidar,'' \emph{Black Hat Europe},
  vol.~11, no. 2015, p. 995, 2015.

\bibitem{cao2019adversarial}
Y.~Cao, C.~Xiao, B.~Cyr, Y.~Zhou, W.~Park, S.~Rampazzi, Q.~A. Chen, K.~Fu, and
  Z.~M. Mao, ``Adversarial sensor attack on lidar-based perception in
  autonomous driving,'' in \emph{Proceedings of the 2019 ACM SIGSAC conference
  on computer and communications security}, 2019, pp. 2267--2281.

\bibitem{spina2021integrating}
M.~G. Spina, S.~Gualtieri, and F.~De~Rango, ``Integrating ros and gazebo tools
  with a network security module to support secure autonomous robot
  coordination.'' in \emph{SIMULTECH}, 2021, pp. 369--377.

\bibitem{dieber2017security}
B.~Dieber, B.~Breiling, S.~Taurer, S.~Kacianka, S.~Rass, and P.~Schartner,
  ``Security for the robot operating system,'' \emph{Robotics and Autonomous
  Systems}, vol.~98, pp. 192--203, 2017.

\bibitem{koubaa2019micro}
A.~Koub{\^a}a, A.~Allouch, M.~Alajlan, Y.~Javed, A.~Belghith, and M.~Khalgui,
  ``Micro air vehicle link (mavlink) in a nutshell: A survey,'' \emph{IEEE
  Access}, vol.~7, pp. 87\,658--87\,680, 2019.

\bibitem{kwon2018empirical}
Y.-M. Kwon, J.~Yu, B.-M. Cho, Y.~Eun, and K.-J. Park, ``Empirical analysis of
  mavlink protocol vulnerability for attacking unmanned aerial vehicles,''
  \emph{IEEE Access}, vol.~6, pp. 43\,203--43\,212, 2018.

\bibitem{Obstacle2005Ribeiro}
M.~I. Ribeiro, ``Obstacle avoidance,'' \emph{The Robotics WEBook}, 2005.

\bibitem{Realtime1995Khatib}
O.~Khatib, ``Real-time obstacle avoidance for manipulators and mobile robots,''
  in \emph{1985 IEEE International Conference on Robotics and Automation},
  vol.~2, 1985, pp. 500 -- 505.

\bibitem{Geometry2004Tabak}
J.~Tabak, \emph{Geometry: The Language of Space and Form}.\hskip 1em plus 0.5em
  minus 0.4em\relax New York, USA: Infobase Publishing, 2004.

\bibitem{rozenbeek2020evaluation}
D.~J. Rozenbeek, ``Evaluation of drone neutralization methods using radio
  jamming and spoofing techniques,'' {PhD} dissertation, School of Electrical
  Engineering and Computer Science. Stockholm, Sweden, 2020.

\bibitem{Drone2018Pokhrel}
\BIBentryALTinterwordspacing
N.~Pokhrel, ``\BIBforeignlanguage{English}{{Drone Obstacle Avoidance and
  Navigation Using Artificial Intelligence}},'' Master's thesis, Aalto
  University. School of Science, 2018. [Online]. Available:
  \url{http://urn.fi/URN:NBN:fi:aalto-201806012988}
\BIBentrySTDinterwordspacing

\end{thebibliography}
